# Efficient Design of Reversible Sequential Circuit

Md. Selim Al Mamun[1], Indrani Mandal[2], Md. Hasanuzzaman[3]
[1, 2](Computer Science and Engineering, Jatiya Kabi Kazi Nazrul Islam University, Bangladesh)
[3](Computer Science and Engineering, Dhaka University, Bangladesh)

***ABSTRACT:*** *Reversible logic has come to the forefront of theoretical and applied research today. Although many researchers are investigating techniques to synthesize reversible combinational logic, there is little work in the area of sequential reversible logic. Latches and flip-flops are the most significant memory elements for the forthcoming sequential memory elements. In this paper, we proposed two new reversible logic gates MG-1 and MG-2. We then proposed new design techniques for latches and flip-flops with the help of the new proposed gates. The proposed designs are better than the existing ones in terms of number of gates, garbage outputs and delay.*

## I. INTRODUCTION

In recent year reversible logic has been considered as an important issue for computer design. The primary reason for this is the increasing demands for lower power devices. It has been proved that for irreversible logic computations, each bit of information loss generates kTln2 joules of heat energy, where k is Boltzmann's constant and T is the absolute temperature at which computation is performed [1]. The generated heat can be problematic for larger circuits. Bennett [2] showed that circuits must be built using reversible logic gates only to avoid the heat dissipation.

Reversible logic has a one to one correspondence between its inputs and outputs. Thus reversible computing does not result in information loss during the computation process. Hence it naturally takes care of heating generated due to information loss. Despite the great potential of reversible logic and these endorsements from the leaders in the field, a little work has been done in the area of sequential reversible logic. In this paper, we studied the existing works in this area and proposed some new designs for sequential circuits - D latch and JK latch.

Many researchers are working in the field of reversible logic to optimize the circuit design. The main challenges of designing reversible circuits are to reduce the number of gates, garbage outputs, constant inputs, propagation delay and quantum cost. While designing the proposed latches, we tried to optimize the number of gates, number of garbage outputs, delay and hardware complexity. In this paper, two new reversible logic gates 'Mamun Gate' MG-1 and MG-2 are proposed for the implementation of D latch and JK latch.

The rest of the paper is organized as follows: Section 2 provides the readers with the necessary background in reversible logic. Section 3 presents our proposed reversible gate MG1 and MG2. Section 4 describes proposed design and synthesis of reversible sequential circuit. Section 5 presents the comparisons of our design with the existing ones in literature and finally this paper is concluded with section 6

## II. REVERSIBLE LOGIC GATES

In this section, some basic definitions related to reversible logic are presented. We formally define reversible gate, garbage output, then we present some popular reversible gates along with their input-output specifications. We also describe the quantum equivalent circuits for the popular reversible gates.

**Definition 2.1.** A Reversible Gate is a k-input, k-output (denoted by k*k) circuit that produces a unique output pattern for each possible input pattern [3].

**Definition 2.2.** Unwanted or unused outputs which are needed to maintain reversibility of a reversible gate (or circuit) are known as Garbage Outputs. The garbage output of Feynman gate [4] is shown in Fig.1 with *.

**Definition 2.3.** The delay of a logic circuit is the maximum number of gates in a path from any input line to any output line. This definition is based on the assumptions that: (i) Each gate performs computation in one unit time and (ii) All inputs to the circuit are available before the computation begins.

**Definition 2.4.** The input vector, $I_v$ and output vector, $O_v$ for 2*2 Feynman Gate (FG) is defined as follows: $I_v = (A, B)$ and $O_v = (P = A, Q = A \oplus B)$. The block diagram and quantum representation for 2*2 Feynman gate is shown in Fig. 1.

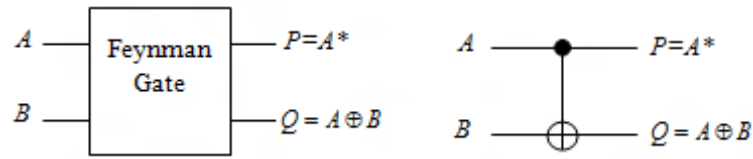

**Figure 1:** Block diagram and quantum representation of a 2*2 Feynman Gate.

**Definition 2.5.** The input vector, $I_v$ and output vector, $O_v$ for 3*3 Toffoli gate (TG) [5] is defined as follows: $I_v = (A, B, C)$ and $O_v = (P = A, Q = B, R = AB \oplus C)$. The block diagram and quantum representation for 3*3 Toffoli gate is shown in Fig. 2.

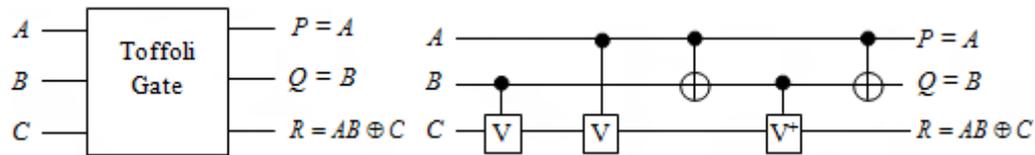

**Figure 2:** Block diagram and quantum representation of a 3*3 Toffoli gate

**Definition 2.6.** The input vector, $I_v$ and output vector, $O_v$ for 3*3 Fredkin gate (FRG) [6] is defined as follows: $I_v = (A, B, C)$ and $O_v = (P = A, Q = \overline{A}B \oplus AC, R = \overline{A}C \oplus AB)$. The block diagram and quantum representation for 3*3 Fredkin gate is shown in Fig. 3.

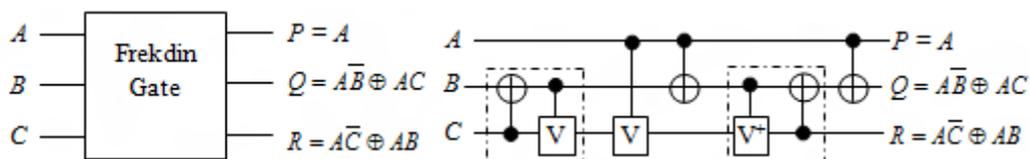

**Figure 3:** Block diagram and quantum representation of a 3*3 Fredkin gate

**Definition 2.7.** The input vector, $I_v$ and output vector, $O_v$ for 3*3 *Peres gate* (PG)[7] is defined as follows: $I_v = (A, B, C)$ and $O_v = (P = A, Q = A \oplus B, R = AB \oplus C)$. The block diagram and quantum representation for 3*3 Peres gate is shown in Fig. 4.

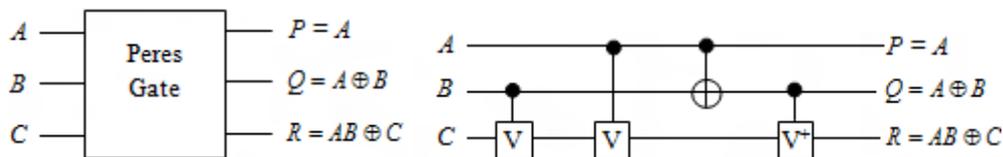

**Figure 4:** Block diagram and quantum representation of a 3*3 Peres gate

### III. PROPOSED REVERSIBLE LOGIC GATES

In this paper we proposed two new 4x4 reversible gates named 'Mamun Gate': MG-1 and MG-2 Here in this section we described the input-output specifications, block diagrams and truth tables of the proposed gates.

#### 3.1. Proposed MG-1 Gate

The input and output vectors of MG-1 gate are $I_v=(A,B,C,D)$ and $O_v=(P = A \oplus D, Q = \overline{A}B \oplus A\overline{C}, R = \overline{A}C \oplus AB, S = \overline{A}C \oplus AB \oplus D)$. The block diagram and truth table of a 4x4 MG-1 are given in Fig. 5 and Table I respectively.

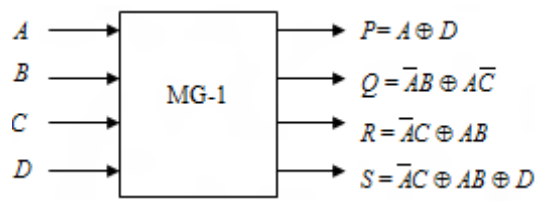

Figure 5: Block diagram of a 4x4 MG-1 Gates

Table I: Truth Table of a 4x4 MG-1 gate

| A | B | C | D | $P = A \oplus D$ | $Q = \overline{A}B \oplus A\overline{C}$ | $R = \overline{A}C \oplus AB$ | $S = \overline{A}C \oplus AB \oplus D$ |
|---|---|---|---|---|---|---|---|
| 0 | 0 | 0 | 0 | 0 | 0 | 0 | 0 |
| 0 | 0 | 0 | 1 | 1 | 0 | 0 | 1 |
| 0 | 0 | 1 | 0 | 0 | 0 | 1 | 1 |
| 0 | 0 | 1 | 1 | 1 | 0 | 1 | 0 |
| 0 | 1 | 0 | 0 | 0 | 1 | 0 | 0 |
| 0 | 1 | 0 | 1 | 1 | 1 | 0 | 1 |
| 0 | 1 | 1 | 0 | 0 | 1 | 1 | 1 |
| 0 | 1 | 1 | 1 | 1 | 1 | 1 | 0 |
| 1 | 0 | 0 | 0 | 1 | 1 | 0 | 0 |
| 1 | 0 | 0 | 1 | 0 | 1 | 0 | 1 |
| 1 | 0 | 1 | 0 | 1 | 0 | 0 | 0 |
| 1 | 0 | 1 | 1 | 0 | 0 | 0 | 1 |
| 1 | 1 | 0 | 0 | 1 | 1 | 1 | 1 |
| 1 | 1 | 0 | 1 | 0 | 1 | 1 | 0 |
| 1 | 1 | 1 | 0 | 1 | 0 | 1 | 1 |
| 1 | 1 | 1 | 1 | 0 | 0 | 1 | 0 |

### 3.2. Proposed MG-2 Gate

The input and output vectors of MG-2 gate are $I_v$ = (A, B, C, D) and $O_v$ (P=A, $Q = \overline{A}B \oplus A\overline{C}$, $R = \overline{A}C \oplus AB$, $S = A \oplus D$. The block diagram and truth table of a 4x4 MG-2 are given in Fig. 6 and Table II respectively..

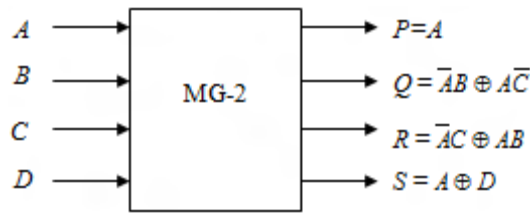

Figure 2: Block diagram of a 4x4 MG-2 Gate

Table II: Truth table of a 4x4 MG-2 gate

| A | B | C | D | $P = A$ | $Q = \overline{A}B \oplus A\overline{C}$ | $R = \overline{A}C \oplus AB$ | $S = A \oplus D$ |
|---|---|---|---|---|---|---|---|
| 0 | 0 | 0 | 0 | 0 | 0 | 0 | 0 |
| 0 | 0 | 0 | 1 | 0 | 0 | 0 | 1 |
| 0 | 0 | 1 | 0 | 0 | 0 | 1 | 0 |
| 0 | 0 | 1 | 1 | 0 | 0 | 1 | 1 |
| 0 | 1 | 0 | 0 | 0 | 1 | 0 | 0 |
| 0 | 1 | 0 | 1 | 0 | 1 | 0 | 1 |
| 0 | 1 | 1 | 0 | 0 | 1 | 1 | 0 |
| 0 | 1 | 1 | 1 | 0 | 1 | 1 | 1 |
| 1 | 0 | 0 | 0 | 1 | 1 | 0 | 1 |
| 1 | 0 | 0 | 1 | 1 | 1 | 0 | 0 |
| 1 | 0 | 1 | 0 | 1 | 0 | 0 | 1 |
| 1 | 0 | 1 | 1 | 1 | 0 | 0 | 0 |
| 1 | 1 | 0 | 0 | 1 | 1 | 1 | 1 |
| 1 | 1 | 0 | 1 | 1 | 1 | 1 | 0 |
| 1 | 1 | 1 | 0 | 1 | 0 | 1 | 1 |
| 1 | 1 | 1 | 1 | 1 | 0 | 1 | 0 |

## IV. DESIGN OF REVERSIBLE LATCHES USING MG GATES

Many researchers proposed their own design for sequential circuit. The designs vary in terms of number of gates and quantum costs. In this section we described our proposed design of reversible D-Latch and JK Latch.

### 4.1. Proposed D Latch
The D Latch can be realized by only one MG-1 gate. The characteristic equation of D latch is $Q = CLK.D \oplus \overline{CLK}.Q$. The D Latch can be mapped with MG-1 by giving *CLK*, D, Q and 1 respectively in 1st, 2nd, 3rd and 4th inputs of MG-1. The Fig. 7 shows the proposed D latch.

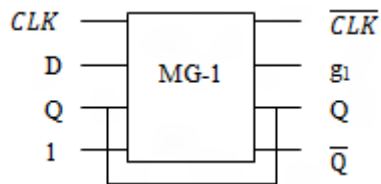

Figure 3: Proposed design of D Latch

The Master-Slave D flip-flop can be realized by two MG-1 gates. No extra gate is needed to generate $\overline{CLK}$. The realization of Master-Slave D flip-flop is shown in Fig. 8.

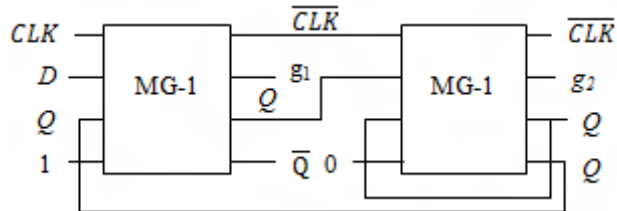

Figure 4: Master-Slave D flip-flop

### 4.2. Proposed JK Latch
The JK latch can be realized by one MG-1 gate and one MG-2 gate. The characteristic equation of JK latch is $Q = \overline{CLK}.Q \oplus CLK(J\overline{Q} \oplus \overline{K}Q)$. The JK latch can be mapped first giving inputs *Q, J, K, 0*, respectively in 1st, 2nd, 3rd and 4th inputs of MG-2 gate. Then *CLK*, $J\overline{Q} \oplus \overline{K}Q$ (2nd output of MG-2 gate), *Q and 1* respectively in 1st, 2nd, 3rd and 4th inputs of MG-1. The proposed JK flip-flop is shown in Fig. 9.

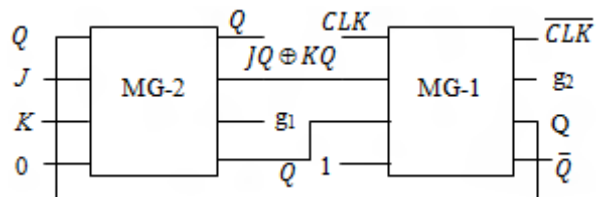

Figure 5: Proposed JK Flip-Flop

The Master-Slave JK flip-flop can be realized by two MG-1 gates and one MG-2 gate. The proposed Master-Slave JK flip-flop is shown in Fig. 10.

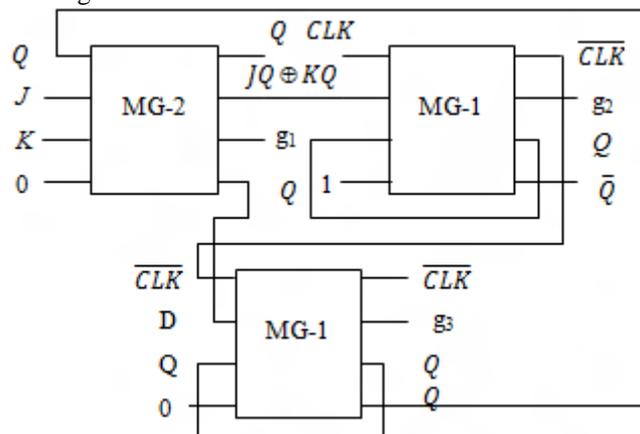

Figure 6: Master-Slave JK flip-flop.

*Efficient Design of Reversible Sequential Circuit*

## V. RESULTS AND DISCUSSIONS

### 5.1. Evaluation of Proposed Reversible MG gates
The proposed reversible gates are 1 through 4x4 gates. With the help of proposed gates design of different latches and flip-flops has been improved. As latches are most important memory block used in RAM and FPGA, this gate can contribute significantly in reversible community

### 5.2. Evaluation of Proposed D Latch
Both the designs of D latch and Master-Slave D latch are optimized than the existing designs. The comparisons are shown in Tables III and IV.

Table III: Comparisons of different designs of D latches

| D Latch design | Cost | | |
| --- | --- | --- | --- |
| | No. of gates | Garbage Outputs | Delay |
| Proposed | 1 | 1 | 1 |
| Existing [8] | 5 | 6 | 5 |
| Existing [9] | 2 | 2 | 2 |
| Existing [10] | 1 | 2 | 1 |
| Existing [11] | 5 | 5 | 5 |
| Existing [12] | 3 | 4 | 3 |
| Existing [13] | 1 | 1 | 1 |

Table IV: Comparisons of different designs of Master-Slave D latches

| Master Slave D Latch | Cost | | |
| --- | --- | --- | --- |
| | No. of gates | Garbage Outputs | Delay |
| Proposed | 2 | 2 | 2 |
| Existing [8] | 11 | 12 | 5 |
| Existing [9] | 5 | 4 | 2 |
| Existing [13] | 3 | 2 | 3 |

From the comparison tables we can see that the proposed designs of D latch are better than existing ones in terms of number of gates, number of garbage outputs and delay.

### 5.3. Evaluation of Proposed JK Latch
Both of our proposed design of JK latch and Master-Slave JK latch are better than the existing ones in literature. Tables V and VI show the comparisons of different design of JK latches.

Table V: Comparisons of different designs of JK latches

| JK Latch design | Cost | | |
| --- | --- | --- | --- |
| | No. of gates | Garbage Outputs | Delay |
| Proposed | 2 | 2 | 2 |
| Existing [8] | 4 | 8 | 4 |
| Existing [9] | 3 | 3 | 3 |
| Existing [10] | 2 | 3 | 2 |
| Existing [13] | 2 | 2 | 2 |

Table VI: Comparisons of different designs of JK latches

| Master-Slave JK Latch | Cost | | |
| --- | --- | --- | --- |
| | No. of gates | Garbage Outputs | Delay |
| Proposed | 3 | 3 | 3 |
| Existing [8] | 12 | 14 | 12 |
| Existing [9] | 6 | 5 | 6 |
| Existing [13] | 3 | 3 | 3 |

From the comparison tables we can see that the proposed designs of JK latches are better than existing ones in terms of number of gates, number of garbage outputs and delay.

## VI. CONCLUSION

Reversible latches are going to be the main memory block for the forthcoming quantum devices. In this paper we proposed optimized reversible D latch and JK latches with the help of proposed 'Mamun' gates. Appropriate algorithms and theorems are presented to clarify the proposed design and to establish its efficiency. We compare our design with the existing designs available in literature which claims our success in terms of number of gates, number of garbage outputs and delay. This optimization can contribute significantly in reversible logic community.